\newcommand{\NPA}[3]{Nucl.\ Phys.\ A\ {\bf #1},\ #2 (#3)}

\newcommand{\diracslash}[1]{#1\llap{/\kern2pt}}

\newcommand{\be}{\begin{equation}}
\newcommand{\ee}{\end{equation}}
\newcommand{\bea}{\begin{eqnarray}}
\newcommand{\eea}{\end{eqnarray}}
\newcommand{\ba}[1]{\begin{array}{#1}}
\newcommand{\ea}{\end{array}}

\documentclass[prd,aps,floats,nofootinbib,tightenlines,showpacs]{revtex4}
\usepackage{epsfig,graphicx,pstricks}
\usepackage{psfrag}
\usepackage{color}
\usepackage{amsmath}
\usepackage{amsfonts}
\usepackage{amssymb}
\usepackage{textcomp}
\usepackage{multirow}
\usepackage{subfigure}

\begin{document}

\title {Transport properties of hadronic matter in magnetic field}
\author{Guru Kadam }
\email{guruprasad@prl.res.in}
\affiliation{Theory Division, Physical Research Laboratory,
Navrangpura, Ahmedabad 380 009, India}

\date{\today} 

\def\be{\begin{equation}}
\def\ee{\end{equation}}
\def\bearr{\begin{eqnarray}}
\def\eearr{\end{eqnarray}}
\def\zbf#1{{\bf {#1}}}
\def\bfm#1{\mbox{\boldmath $#1$}}
\def\hf{\frac{1}{2}}
\def\sl{\hspace{-0.15cm}/}
\def\omit#1{_{\!\rlap{$\scriptscriptstyle \backslash$}
{\scriptscriptstyle #1}}}
\def\vec#1{\mathchoice
        {\mbox{\boldmath $#1$}}
        {\mbox{\boldmath $#1$}}
        {\mbox{\boldmath $\scriptstyle #1$}}
        {\mbox{\boldmath $\scriptscriptstyle #1$}}
}

\begin{abstract}
We study the effect of magnetic field  on the transport properties like shear and bulk viscosities of hot and dense hadronic matter within hadron resonance gas model. We estimate the bulk viscosity using low energy theorems for bilocal correlators of the energy momentum tensor generalized to finite temperature, density and magnetic field. We use Gaussian ansatz for the spectral function at low frequency. We estimate shear viscosity coefficient using molecular kinetic theory. We find that vacuum contribution due to finite magnetic field dominates the bulk viscosity ($\zeta$) for the temperatures up to $0.1 GeV$ and increases with magnetic field while ratio $\zeta/s$ decreases with magnetic field. We also find that shear viscosity coefficient of hadronic matter decreases with magnetic field. 
\end{abstract}


\maketitle

\section{Introduction}
Study of physical system under the magnetic field is an interesting as well as important problem in its own right. Magnetic field is such a ubiquitous physical quantity that it exist at all levels, from the microscopic subnuclear scale to the macroscopic scale of galaxies and clusters of galaxies. Matter in the universe exist in various phases and this phase structure depends on microscopic interaction between constituents of matter and external macroscopic parameters.  Quantum chromodynamics which is the theory of strong interaction predicts various phases of strongly interacting matter. At low baryon density there exist at least two phases of QCD viz: hadronic phase which exist at low temperature and quark gluon plasma (QGP) phase which exist at very high temperature. Standard big bang theory predicts the phase transition from QGP to hadronic phase as universe expanded and cooled at about $10^{-5}$ seconds after the big bang. Little bang experiments at RHIC and LHC have also observed such transition. Besides temperature $(T)$ and baryon chemical potential $(\mu_{b})$, magnetic field $(B)$ is another important macroscopic parameter which can affect the QCD phase diagram. One of the important phenomena exhibited by strongly interacting matter in magnetic field is the magnetic catalysis effect where the dynamical chiral symmetry breaking is enhanced due to increase in $\langle \bar qq\rangle$ condensates whence the dynamical mass with magnetic field\cite{shovkovy,delia,partyka}. However, the opposite effect has also been found in various effective models\cite{alaya,ferreira} as well as in lattice QCD\cite{balicm} where the dynamical mass decreases with magnetic field. This is called inverse magnetic catalysis effect.  Another interesting  phenomenon that has attracted attention in the context of off central relativistic heavy ion collision is the chiral magnetic effect\cite{FukuKharzeev}. In presence of CP violation, even if locally, left handed and right handed quarks generate a current in the direction of magnetic field leading to the separation of charges.

There have been wide range of physical systems where magnetic field exist and can have drastic impact on aspects of phase transition and equation of state. For instance, in non-central heavy ion collision external magnetic field is generated by spectators and its strength reaches up to masses of hadrons. Neutron stars are natural laboratories where it is expected that very strong magnetic field can exist. In such a cases study of QCD matter under magnetic field becomes very important.

Study of transport properties is an important aspect of strongly interacting matter. When the system is perturbed from its equilibrium, transport properties like shear viscosity $(\eta)$, bulk viscosity $(\zeta)$ and thermal conductivity $(\sigma)$ governs its decay to equilibrium state. In the relativistic heavy ion collision where spatial anisotropy is converted into momentum anisotropy during hydrodynamical evolution, shear viscosity coefficient govern the equilibration of this momentum anisotropy. It has been predicted theoretically by ADS/CFT duality that $\eta/s$ for most ideal fluid cannot be lower than $1/4\pi$\cite{kss}. It has been actually observed at RHIC and LHC that QGP produced in heavy ion collision behaves as perfect fluid and have small value of shear viscosity\cite{roma,hirano}. Another important transport coefficient which plays crucial role in hydrodynamical evolution of strongly interacting matter especially around QCD phase transition  is bulk viscosity. There has been various attempts to estimate bulk viscosity of QCD matter and rise in ratio $\zeta/s$ has been observed for both hadronic phase\cite{guru} and quark-gluon matter phase\cite{kharzeev}. 

Presence of magnetic field breaks rotational $SO(3)$ symmetry. This leads to anisotropy in transport properties of strongly interacting matter. In this regards, there have been very interesting study on transport properties of QCD matter using holographic correspondence\cite{critelli}, Kubo formalism\cite{huang} and anisotropic hydrodynamics\cite{huangrishke}. Study of QCD equation of state in external magnetic field is already in literature\cite{bali}. But the transport properties of hadronic matter in magnetic field has been rarely studied. In present study we calculate both shear and bulk viscosity coefficients of hot and dense hadronic matter under background magnetic field. We use low energy theorems of QCD for bilocal correlator of trace of energy momentum tensor generalized to finite temperature ($T$), chemical potential ($\mu$) and magnetic field ($B$) to estimate bulk viscosity. For this purpose we use Gaussian ansatz for spectral density. We estimate the shear viscosity coefficient using molecular kinetic theory within excluded volume hadron resonance gas model.   
 
We organize our paper as follows. In next section we derive expression for bulk viscosity starting from Kubo's formula. We use low energy theorems of QCD generalized at finite temperature, density and magnetic field to obtain sum rule. To extract bulk viscosity using this sum rule we make Gaussian ansatz for spectral function at low frequency. Then we use molecular kinetic theory to obtain shear viscosity coefficient within excluded volume hadron resonance gas model. In section III we recapitulate the results of HRG in background magnetic field. In section IV we discuss the results and finally in section V we summarize and make conclusion.

\section{Transport properties in magnetic field}
\subsection{Bulk viscosity}
Using the fact that physical pressure is invariant under renormalization group transformation, low-energy theorems (LET) of QCD can be derived at finite temperature $(T)$, quark chemical potential $(\mu_{q})$ and magnetic field ($B$)\cite{agasian} where the bilocal correlators can be expressed in terms of derivatives with respect to these physical parameters and renormalized quark masses $m_{q}$. For any operator $\hat O$ of canonical dimension $d$ constructed from quark or gluon fields, they are written as

\be
\bigg(T \frac{\partial}{\partial T}+\sum_{q}\mu_{q}\frac{\partial}{\partial \mu_{q}}+2B\frac{\partial}{\partial B}-d\bigg)^{n}\langle \hat O \rangle=\int d^{4}x_{n}...\int d^{4}x_{1}\langle \theta^{g}_{\mu\mu}(x_{n})...\theta^{g}_{\mu\mu}(x_{1})\hat O(0)\rangle
\label{LET}
\ee
 
This is general relation valid at all $T$, $\mu$ and $B$ provided $\lambda_{0} $ $\gg$ $T$, $\mu$, $B$, $\lambda $. Here $\lambda $ is dimensionfull parameter in the theory and $\lambda_{0} $ is the scale at which UV divergences are regularized. This fact is necessary to ensure that electromagnetic corrections do not appear in $\beta$-function. Anomalous dimension for $\hat O$ do not appear in these relations since we restrict to lowest order of expansion in $\beta$-function.

For $n=1$, LET for gluon and quark fields can be written as

\be
\int d^{4}x \langle \theta^{g}_{\mu\mu}(x)\theta^{g}_{\mu\mu}(0)\rangle=(\hat D-4)\langle \theta^{g}_{\mu\mu} \rangle
\label{LET1}
\ee
\be
\int d^{4}x \langle \theta^{g}_{\mu\mu}(x)\theta^{q}_{\mu\mu}(0)\rangle=(\hat D-3)\langle \theta^{q}_{\mu\mu} \rangle
\label{LET2}
\ee
where $\hat D\equiv T \frac{\partial}{\partial T}+\sum_{q}\mu_{q}\frac{\partial}{\partial \mu_{q}}+2B\frac{\partial}{\partial B}$.

Kubo's formula\cite{kubo} expresses the linear response of a system to external time dependent perturbation. In the static limit of the correlation function of the trace of the stress tensor, Kubo's formula for bulk viscosity can be written as\cite{kharzeev}

\be
\zeta=\frac{1}{9}\lim_{\omega\rightarrow 0}\frac{1}{\omega}\int_{0}^{\infty}dt\int d^{3}r e^{\iota\omega t}\langle[\theta_{\mu\mu}(x),\theta_{\mu\mu}(0)]\rangle
\label{bv}
\ee

Defining spectral density as $\rho(\omega,\boldsymbol p)=-\frac{1}{\pi}Im G^{R}(\omega,\boldsymbol p)$ one can recast Eq.\ref{bv} in terms of spectral density as
\be
2\int_{0}^{\infty}\frac{\rho(\omega,\boldsymbol{0})}{\omega}d\omega=\int d^{4}x\langle \theta_{\mu\mu}(x)\theta_{\mu\mu}(0)\rangle
\ee

Using Eq.\ref{LET1} and \ref{LET2}, r.h.s of above equation can be written as

\be
\int d^{4}x\langle \theta_{\mu\mu}(x)\theta_{\mu\mu}\rangle \backsimeq (\hat D-4)\langle \theta_{\mu\mu}\rangle+(\hat D-2)\langle \theta_{\mu\mu}^{q}\rangle
\ee

Here we have neglected term proportional to $m_{q}^{2}$.
Since physical pressure is invariant under RG transformations one can show that\cite{agasian}
\be
\langle \theta_{\mu\mu}\rangle=4\varepsilon_{v}+(\hat D-4)P_{*}
\label{theta}
\ee

where $\varepsilon_{v}$ is energy density of QCD vacuum and $P_{*}$ is thermodynamic pressure. 
In presence of magnetic field, $SO(3)$ rotational symmetry is broken and pressure in the direction perpendicular to the magnetic field may be different from that of pressure in longitudinal direction. For the thermodynamic system at finite $T$, $\mu$ and $B$, longitudinal thermodynamic pressure in limit $V\longrightarrow \infty $ can be written in terms of energy density ($\varepsilon$), magnetization ($M$), baryon density ($\varrho_{b}$) and entropy density ($s$) as
\be
P_{*}=Ts+BM+\mu \varrho_{b}-\varepsilon
\label{thermrel}
\ee

\be
\langle \theta_{\mu\mu}\rangle=4\varepsilon_{v}+(\varepsilon-3P)_{*}+BM_{*}
\ee

Hence bilocal correlator can be expressed in terms of thermodynamical quantities as

\bearr
\int d^{4}x\langle \theta_{\mu\mu}(x)\theta_{\mu\mu}(0)\rangle&=&-16\varepsilon_{v}-2\sum_{q}m_{q}\langle\bar qq\rangle_{0}+\bigg(T \frac{\partial}{\partial T}+\sum_{q}\mu_{q}\frac{\partial}{\partial \mu_{q}}+2B\frac{\partial}{\partial B}-2\bigg)\sum_{q}m_{q}\langle \bar qq\rangle_{*}\nonumber\\&+&Ts\bigg(\frac{1}{C_{s}^{2}}-3\bigg)+\bigg(\sum_{q}\mu_{q}\frac{\partial}{\partial \mu_{q}}-4\bigg)(\varepsilon-3P)_{*}+2B^{2}\chi-4BM\nonumber\\
&+&B\bigg(T \frac{\partial}{\partial T}+\sum_{q}\mu_{q}\frac{\partial}{\partial \mu_{q}}\bigg)M
\label{sum rule}
\eearr

where $\chi=\partial M/\partial B$ is magnetic susceptibility and $C_{s}^{2}=\partial P/\partial \varepsilon|_{B,\mu}$ is sound velocity at constant magnetic field and chemical potential. Note that our sum rule (\ref{sum rule}) differs from that of \cite{agasian} as far as magnetic field part is concerned.

Our task to extract the bulk viscosity reduces to obtain an analytic expression for spectral density. It is customary to make an ansatz for spectral density relevant to physical system under consideration. At low frequencies we make an ansatz of the form

\be
\frac{\rho(\omega,\mathbf{0})}{\omega}=\frac{9\zeta}{\pi}e^{-(\frac{\omega}{\pi\omega_{0}})^{2}}
\label{ansatz}
\ee

Where $\omega_{0}$ is the scale at which perturbation theory is valid. Let us discuss validity of this spectral function. First, it satisfy the definition of bulk viscosity in terms of spectral function
\be
\zeta=\frac{\pi}{9}\lim_{\omega\rightarrow 0}\frac{\rho(\omega,\mathbf{0})}{\omega}
\ee

Secondly, it is odd under parity as required by parity properties of retarded Green's function, whence of spectral function. Apart from this, ansatz (\ref{ansatz}) reduces to Lorentzian form as in ref. \cite{kharzeev} in small frequency limit. Because the large frequency modes are suppressed in Gaussian form of spectral function, $\zeta/s$ will have lower value as compared to that with Lorentz form of spectral function.  

Sum rule in ref. \cite{kharzeev} is similar to (\ref{sum rule}) with $B=0$.  These authors argue that at high frequency spectral function behaves as $\rho\sim \alpha_{s}^{2}\omega^{4}$ as dictated by perturbation theory. Also at $\omega\gg T$ spectral function is independent of $T$ and since perturbative (divergent) contribution has already been subtracted from r.h.s of sum rule, we need not to subtract it from l.h.s. as well. In ref. \cite{huot} authors show that this sum rule is sensitive to numerically important UV tail $\rho\sim \alpha_{s}^{3}T^{4}$ which would affect the analysis close to the phase transition. But in our study where we are working at temperature much away from QCD phase transition temperature (i.e up to $0.15 GeV$) where hadron resonance gas model is valid, we need not to worry about this tail.
Using ansatz for spectral density (\ref{ansatz}) and sum rule (\ref{sum rule}) we get the expression for bulk viscosity as

\bearr
9\sqrt{\pi}\zeta\omega_{0}&=&-16\varepsilon_{v}-2\sum_{q}m_{q}\langle\bar qq\rangle_{0}+\bigg(T \frac{\partial}{\partial T}+\sum_{q}\mu_{q}\frac{\partial}{\partial \mu_{q}}+2B\frac{\partial}{\partial B}-2\bigg)\sum_{q}m_{q}\langle \bar qq\rangle_{*}\nonumber\\&+&Ts(\frac{1}{C_{s}^{2}}-3)+\bigg(\sum_{q}\mu_{q}\frac{\partial}{\partial \mu_{q}}-4\bigg)(\varepsilon-3P)_{*}+2B^{2}\chi-4BM\nonumber\\
&+&B\bigg(T \frac{\partial}{\partial T}+\sum_{q}\mu_{q}\frac{\partial}{\partial \mu_{q}}\bigg)M
\eearr

We will use this expression to estimate the bulk viscosity of hadronic matter at finite $T$, $\mu$ and $B$ using hadron resonance gas model (HRG). Note that this expression has been derived using longitudinal component of the pressure. One can arrive at somewhat different expression using transverse component of pressure\cite{agasian}. Thus the bulk viscosity is anisotropic in external magnetic field which is reflection of the fact that rotational symmetry is broken in magnetic field\cite{huang,huangrishke,ferrer}. 

\subsection{shear viscosity}
In hadron resonance gas model the repulsive interaction between hadrons can be modeled via excluded volume correction\cite{kapustaolive,rischkegorenstein}. Using molecular kinetic theory one can show that shear viscosity of relativistic hadron gas with hard core radius $r$ under external magnetic field $B$ can be written as\cite{gorenstein,greinerprl,greinerprc}

\be
\eta=\frac{5}{64\sqrt{8}r^{2}}\sum_{i}\langle|\boldsymbol p|\rangle\frac{n_{i}}{n}
\ee

where, $\langle|\boldsymbol p|\rangle=\langle\sqrt{p_{z}^{2}+2eB(k+1/2-s_{z})}\rangle$ is thermal average of momentum and  $n_{i}$ being particle density of i-th species of hadrons with $\sum_{i}n_{i}=n$. Since shear viscosity coefficient is independent of pressure, it is isotropic.

\section{Hadron resonance gas model in magnetic field}
The simplest effective model which captures the physics of hadronic phase of QCD is hadron resonance gas model. QCD equation of state at finite magnetic field in HRG has been studied in ref.\cite{endrodi}. We recapitulate the main results which we will use in our calculation.

The central quantity in HRGM is thermodynamic potential (free energy)
\be
\Omega=E-TS-BM-\mu N
\ee
In terms of densities
\be
\frac{\Omega}{V}=\varepsilon-Ts-Bm_{B}-\mu \varrho_{b}
\ee
Where $m_{b}$ is magnetization density. In thermodynamic limit, $V \longrightarrow\infty$, thermodynamic pressure can be written as
\be
P=-\frac{\Omega}{V}=-f=-(f_{vacuum}+f_{thermal})
\ee
Where $f_{vacuum}$ is vacuum contribution ($T,\mu=0, B\neq0$) to free energy. Hence energy density can be written as
\be
\varepsilon=Ts+Bm_{B}+\mu\varrho_{b}-P
\ee
For the ideal gas of hadrons, free energy for charged component of the gas can be written as

\be
f_{c}= \pm \sum_{h}\sum_{s_{z}}\sum_{k=0}^{\infty}\frac{eB}{4\pi^{2}}\int dp_{z}\bigg(\frac{E(p_{z},k,s_{z})}{2}+T\: log(1\pm e^{-E(p_{z},k,s_{z})/T})\bigg)
\label{frenergy1}
\ee

where $E=\sqrt{p_{z}^{2}+m^{2}+2eB(k+1/2-s_{z})}$ is the energy of charged particle moving freely under external magnetic field pointing in $z$ direction.

Free energy for neutral component of the gas is

\be
f_{n}= \pm \sum_{h}\int d^{3}p\bigg(\frac{E_{0}}{2}+T\: log(1\pm e^{-E_{0}/T})\bigg)
\label{frenergy2}
\ee

where $E_{0}=\sqrt{\boldsymbol p^{2}+m^{2}}$.

Once we know the free energy thermodynamical quantities can be evaluated as, $s=\partial P/\partial T$, $m_{B}=\partial P/\partial B$, $C_{s}^{2}=\partial P/\partial \varepsilon|_{B}$.

Vacuum terms in Eq. (\ref{frenergy1}) and (\ref{frenergy2}) are UV divergent and can be regularized by dimensional regularization and renormalization of $B>0$ free energy can be carried out by subtracting $B=0$ part. Renormalized  vacuum free energies for different spin channels in magnetic field are given by\cite{endrodi}
\be
\Delta f^{vac,r}(spin\: 0)=\frac{(eB)^{2}}{8\pi^{2}}\bigg[\varsigma^{'}(-1,x+1/2)+x^{2}/4-\frac{x^{2}}{2} log(x)+\frac{log(x)+1}{24}\bigg]
\ee

\be
\Delta f^{vac,r}(spin\: 1/2)=-\frac{(eB)^{2}}{4\pi^{2}}\bigg[\varsigma^{'}(-1,x)+\frac{x}{2}log(x)+x^{2}/4-\frac{x^{2}}{2} log(x)-\frac{log(x)+1}{12}\bigg]
\ee

\bearr
\Delta f^{vac,r}(spin\: 1)&=&\frac{3(eB)^{2}}{8\pi^{2}}\bigg[\varsigma^{'}(-1,x-1/2)+\frac{1}{3}(x+1/2)log(x+1/2)\nonumber \\
&+&\frac{2}{3}(x-1/2)log(x-1/2)+x^{2}/4-\frac{x^{2}}{2} log(x)-\frac{7(log(x)+1)}{24}\bigg]
\eearr

where $x=\frac{m_{h}^{2}}{2eB}$ and $\varsigma(-1,x)$ is Hurwitz zeta function whose asymptotic ($x\gg1$) expression is given by

\be
\varsigma^{'}(-1,x)=\frac{1}{12}-\frac{x^{2}}{4}-\bigg(\frac{1}{12}-\frac{x}{2}+\frac{x^{2}}{2}\bigg)
\ee

Thus our results are valid  for magnetic fields for which condition $eB\ll m_{\pi}^{2}/2$ is satisfied.

\section{Results and discussion}

Free energy (Eq. \ref{frenergy1}) has contribution coming from individual hadrons. This contribution depends on external parameters ($T,\mu,B$) and internal quantum numbers (mass, spin, charge and gyromagnetic ratio). Since experimental gyromagnetic ratios are known with small error bars only for  lightest hadrons, we take gyromagnetic ratios, $g_{h}=2q_{h}/e$\cite{endrodi}. We take all the hadrons up to $m_{h}=1.2 GeV$. Parameter $\omega_{0}$ have been chosen equal to $1 GeV$ as in ref.\cite{kharzeev}.

\begin{figure}
 \centering
 \subfigure[]{
 \includegraphics[width=.42\columnwidth]{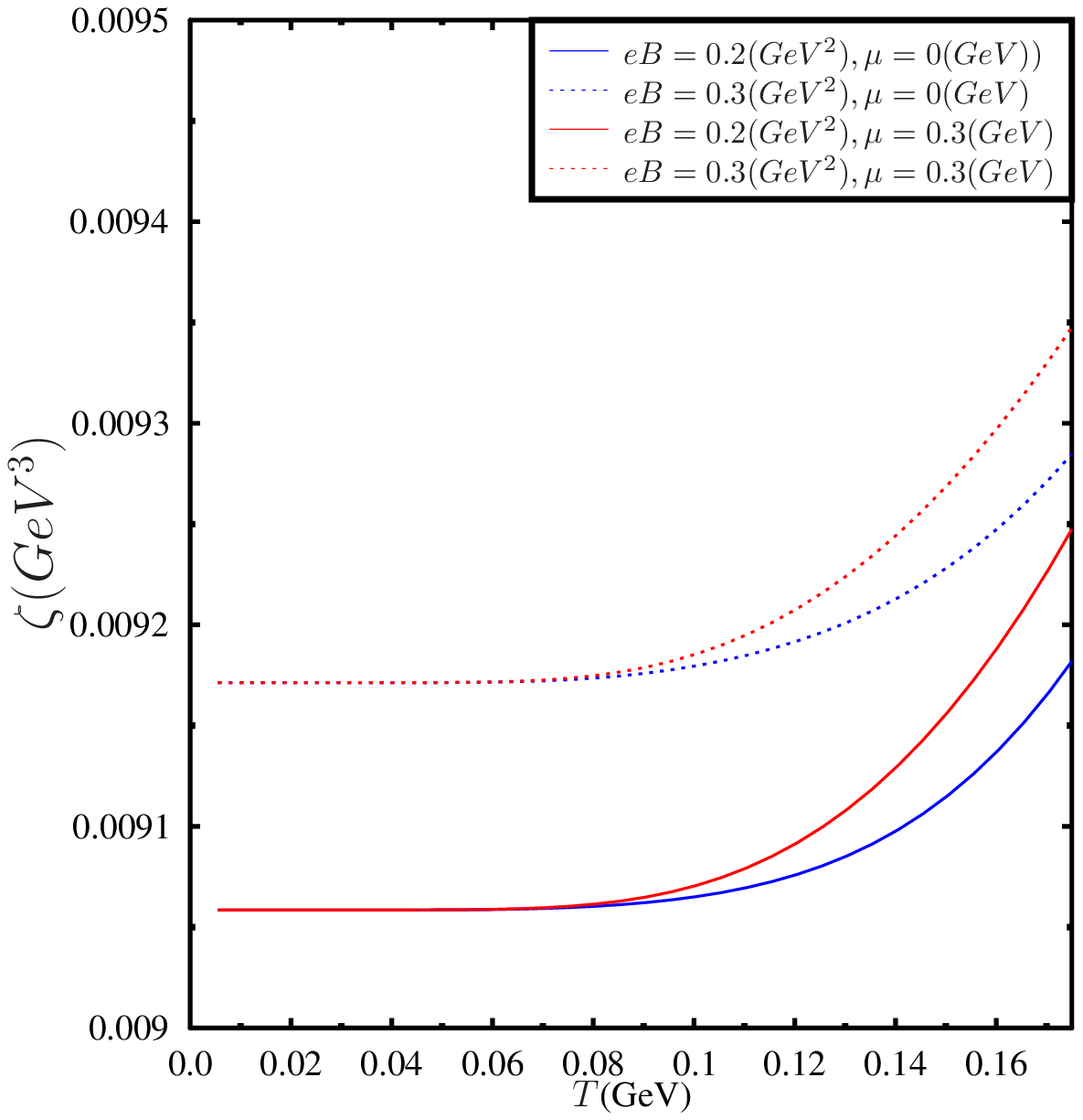}
 }
 \subfigure[]{
 \includegraphics[width=.42\columnwidth]{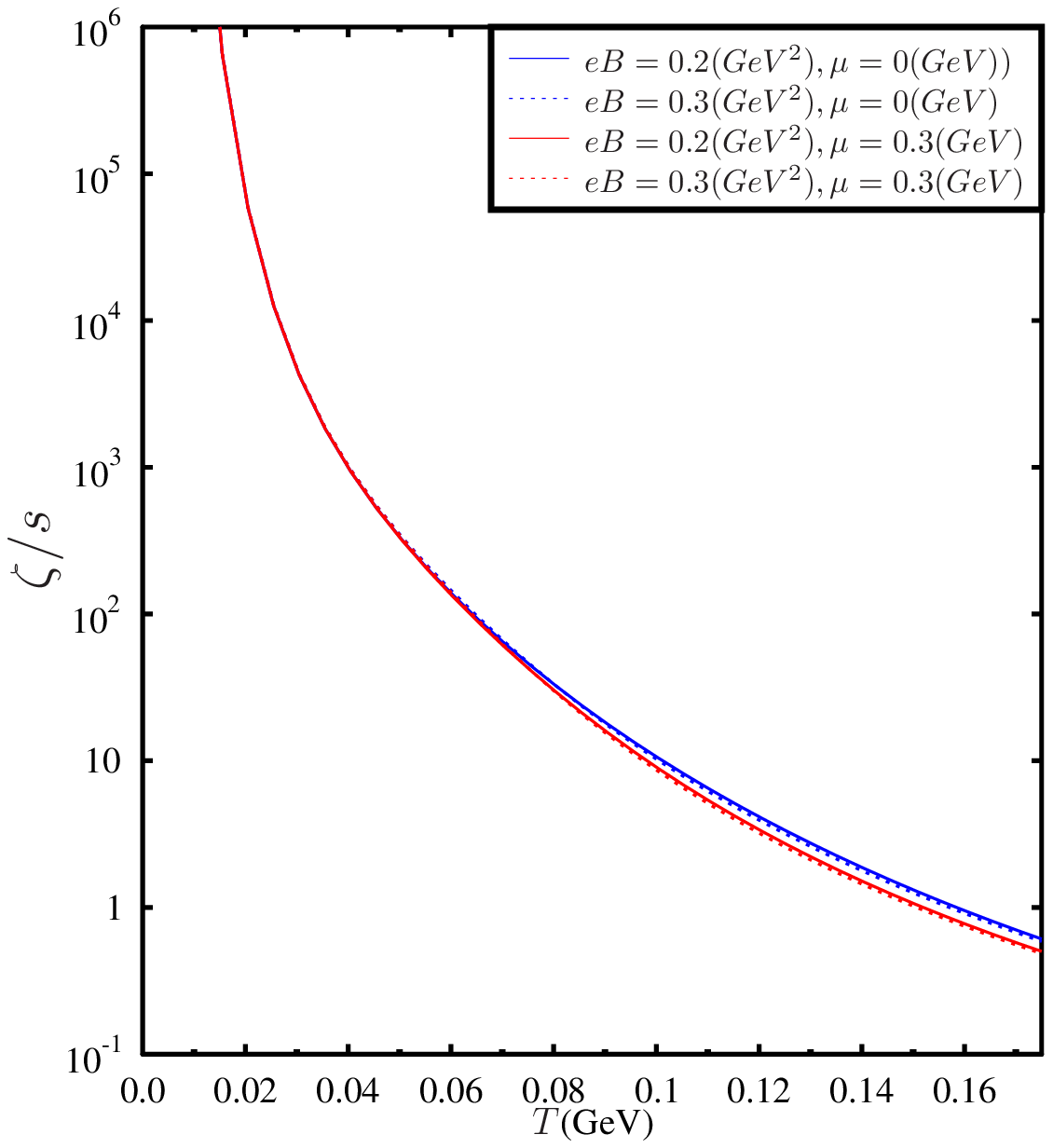}
 }
\caption{ (a)  Bulk viscosity  as a function of temperature. (b) Bulk viscosity in units of entropy density as a function of temperature.
}

\label{zetabis}
 \end{figure}

 \begin{figure}[h]
\vspace{-0.4cm}
\begin{center}
 \includegraphics[width=10cm,height=8cm]{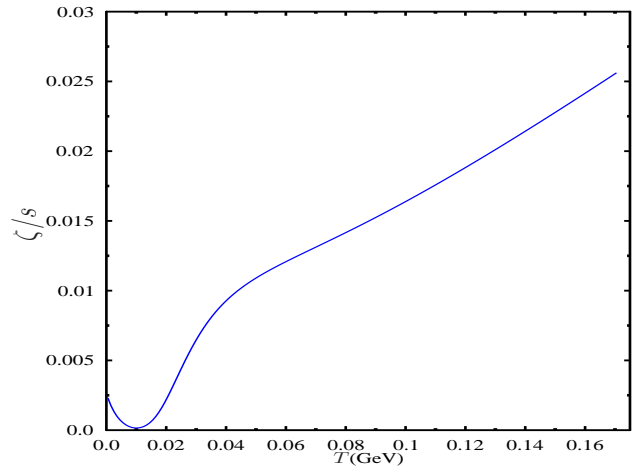}
 
\caption{\label{zetabis1}  Bulk viscosity in units of entropy density as a function of temperature at $eB=0$.
}
\end{center}
 \end{figure}
 
 Fig \ref{zetabis}(a) shows bulk viscosity as a function of temperature at two different magnetic fields and chemical potentials. We note that vacuum contribution due to finite magnetic field dominates the bulk viscosity up to $0.1 GeV$. This behavior may be interpreted as follows. The effective mass of the charged particle in magnetic field is given by
 \be
 m_{*}^{2}=m^{2}+B(1-2s)
 \ee
 where $s$ is total spin of the particle. This effective mass increases with magnetic field for spin $0$ channel but decreases for spin $1$ channel and remains same for spin $1/2$ channel. Thus statistical weight factor, $exp(-\beta m_{*})$ is larger for spin $1$ channel than for spin $0$ channel. At low temperature where the system is dominated by pions, thermal contribution to thermodynamic quantities (pressure, energy density, magnetization, susceptibility) is very small. Hence these quantities are dominated by vacuum part due to finite magnetic field. At finite magnetic field, as the bulk viscosity is proportional to magnetic susceptibility, bulk viscosity has dominant contribution from vacuum susceptibility. Above $T\backsimeq0.1 GeV$, due to thermal excitation of $\rho^{\pm}$ mesons and other heavier hadrons, bulk viscosity rises. Also at finite chemical potential we note that bulk viscosity rises more rapidly as compared to $\mu=0$ case and thermal contribution to the bulk viscosity starts at lower temperature. This is due to thermal excitation of baryons at lower temperature. 
 
 Fig \ref{zetabis}(b) shows bulk viscosity in units of entropy density at finite $\mu$ and $B$. We note that behavior of $\zeta/s$ in magnetic field is opposite to that of $B=0$ case (Fig. \ref{zetabis1}). This is a reflection of the fact that bulk viscosity is non-zero even at $T=0$ while entropy density is zero  so that ratio $\zeta/s$ blows up. As temperature increases, entropy density increases while bulk viscosity remains constant to its vacuum value. Whence $\zeta/s$ decreases.

 \begin{figure}
 \centering
 \subfigure[]{
 \includegraphics[width=.42\columnwidth]{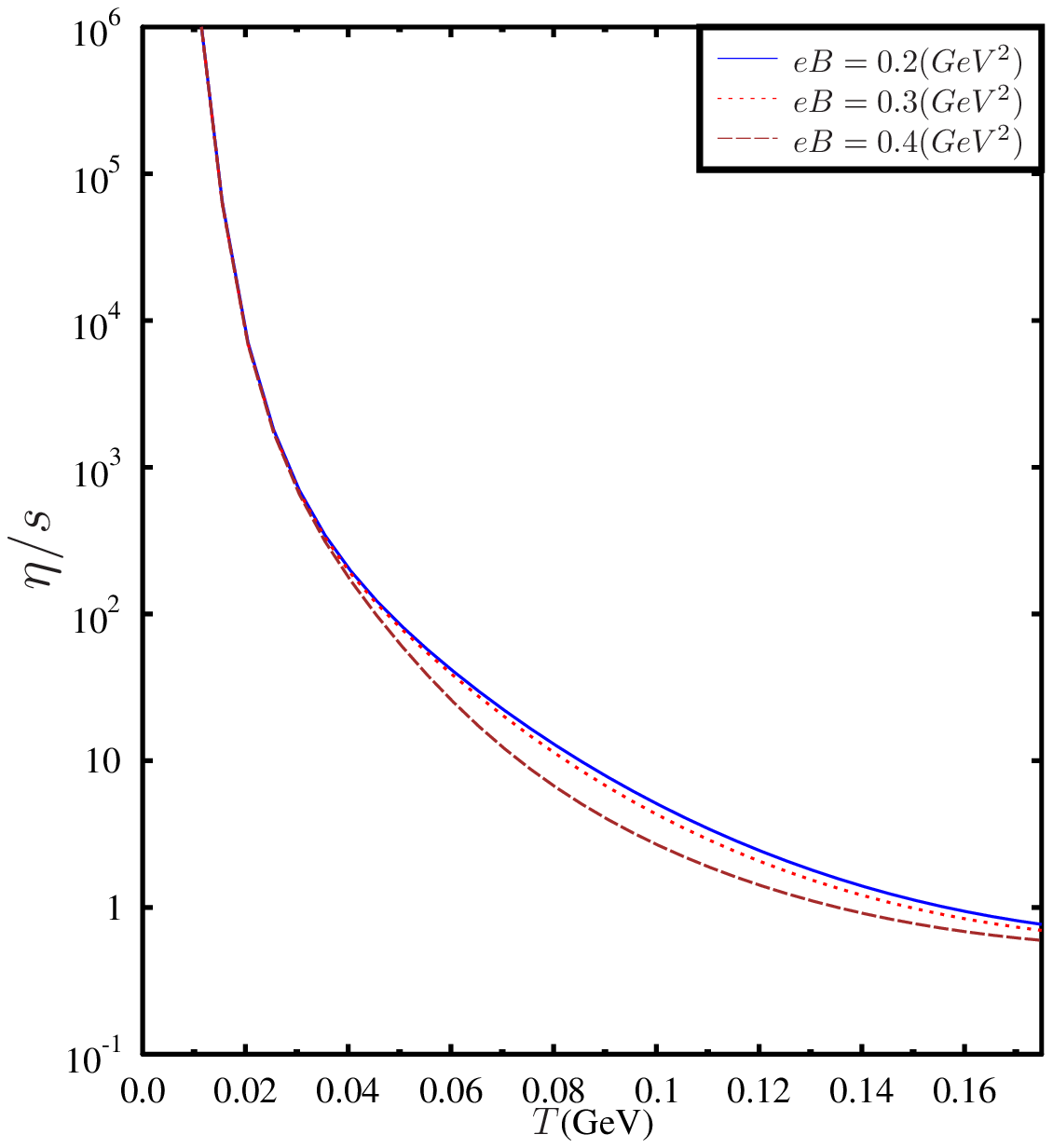}
 }
 \subfigure[]{
 \includegraphics[width=.42\columnwidth]{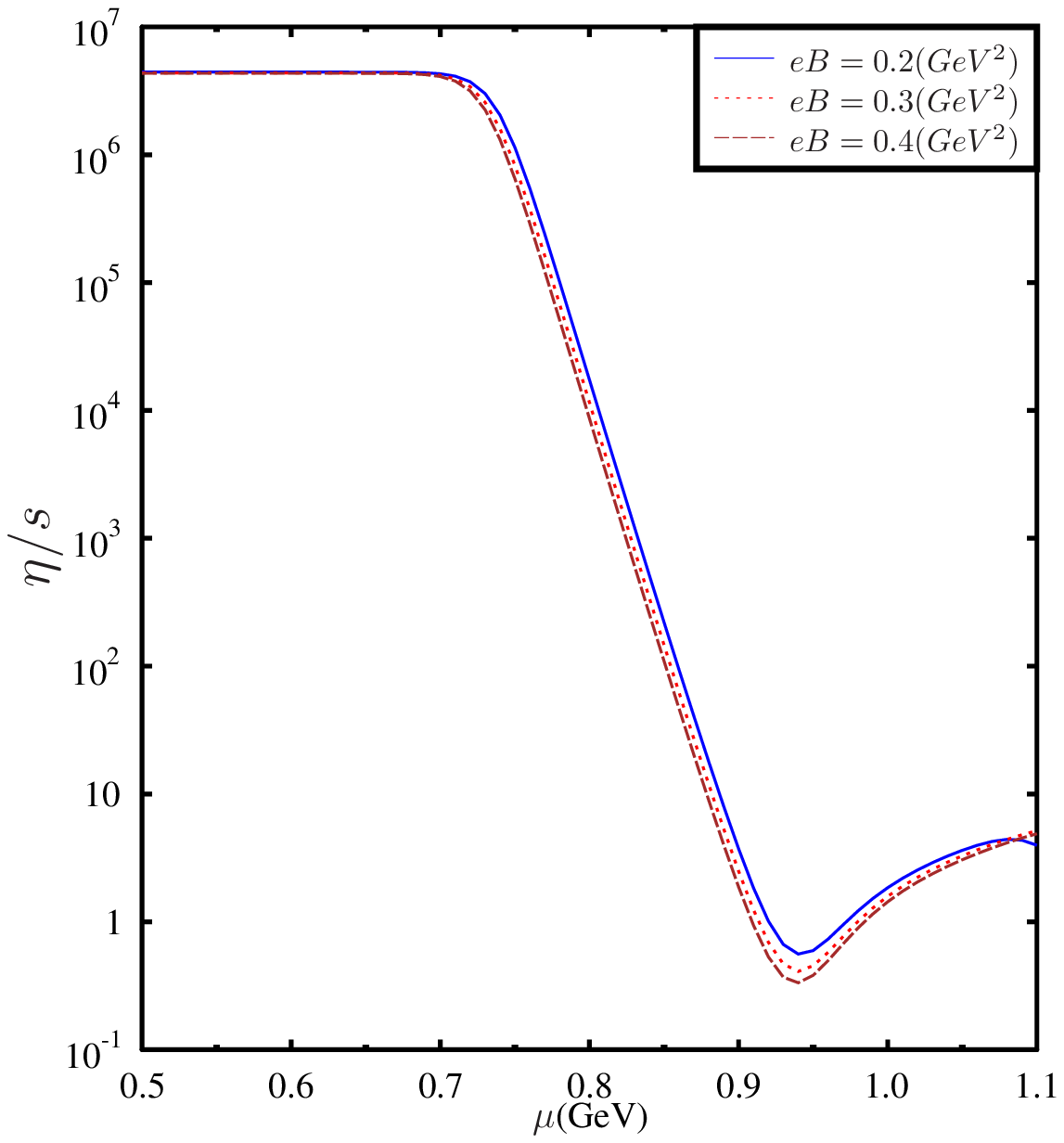}
 }
 \caption{\label{etabis}(a) Shear viscosity in units of entropy density  as a function of temperature at $\mu=0$. (b) $\eta/s$ as a function of baryon chemical potential at $T=5MeV$.}
 \end{figure}

  Fig.\ref{etabis}(a) shows estimate of shear viscosity to entropy ratio  assuming hardcore radii $r_{h}=0.5 fm$ for all hadrons. We note that $\eta/s$  decreases with magnetic field. Such behavior of $\eta/s$ has been observed for quark matter under strong magnetic field\cite{nam} and interpreted as reflection of magnetic catalysis. We also note from Fig.\ref{etabis}(b) that $\zeta/s$ has minimum at $\mu=0.94 GeV$. This minimum has been interpreted as corresponding to nuclear liquid-gas phase transition and has been observed to flattened out above $T=0.025GeV$\cite{guru,itakura,cpsingh}. We observe that this minimum merely shift towards lower value with increasing magnetic field and does not show any flattening.  This may imply that magnetic field does not affect nuclear liquid-gas phase transition.  
 
 \section{summary}
 In this paper we have estimated bulk and shear viscosities of hadronic matter under external magnetic field within hadron resonance gas model. At finite magnetic field vacuum contribution to the bulk viscosity dominates at low temperature which is in contrast to behavior of bulk viscosity at zero magnetic field. This behavior is the reflection of the fact that scalar and vector mesons have different effective mass in magnetic field which affect their thermal  excitation. We mention here that sum rule \ref{sum rule} has certain limitations\cite{rom,huot,moore} and needs to be improved. 
 In shear viscosity channel we found that ratio $\eta/s$ decreases with magnetic field. Also $\eta/s$ shows minimum at $\mu=0.940 GeV$ and this minimum just shift towards smaller value without flattening. This may imply that magnetic field does not affect nuclear liquid-gas phase transition.

 \def\shovkovy{V. Gusynin, V. Miransky, I. Shovkovy, Phys. Rev. Lett. {\bf 73}, 3499 (1994).}
 \def\delia{M. D'Elia, S. Mukherjee and F. Sanfilippo, Phys. Rev. D {\bf 82}, 051501 (2010).}
 \def\partyka{T. L. Partyka, Mod. Phys. Lett. A {\bf 29}, 1450058 (2014).}
 \def\alaya{Alejandro Ayala, M. Loewe, Ana Júlia Mizher, and R. Zamora, Phys. Rev. D {\bf 90}, 036001 (2014).}
 \def\ferreira{M. Ferreira, P. Costa, D. P. Menezes, C. Providencia, and N. N. Scoccola, Phys. Rev. D {\bf 89}, 016002 (2014).}
 \def\kss{P. Kovtun, D.T. Son and A.O. Starinets, Phys. Rev. Lett. {\bf 94}, 111601, (2005).}
 \def\roma{P. Romatschke and U. Romatschke, Phys. Rev. Lett. {\bf 99},172301, (2007).}
 \def\hirano{ T. Hirano and M. Gyulassy, Nucl. Phys. A {\bf  769}, 71, (2006).}
\def\kharzeev{F. Karsch, D. Kharzeev, and K. Tuchin, Phys. Lett. B {\bf 663}, 217 (2008).}
\def\critelli{R. Critelli, S. I. Finazzo, M. Zaniboni, and J. Noronha, Phys.Rev. D {\bf 90}, 066006 (2014)}
\def\huang{Xu-Guang Huang, Dirk H. Rischke, and A. Sedrakian, Annals Phys. {\bf 326}, 3075-3094 (2011)}
\def\huangrishke{Xu-Guang Huang, Mei Huang, Dirk H. Rischke, and A. Sedrakian, Phys.Rev. D {\bf 81}, 045015 (2010)}
\def\bali{G. Bali etal, JHEP 1408, 177 (2014).}
\def\balicm{G. Bali etal, JHEP {\bf 02}, 044 (2012).}
\def\agasian{N. Agasian, JETP Lett. {\bf 95} (2012).}
\def\huot{S. Caron-Huot, Phys. Rev. D {\bf 79}, 125009 (2009).}
\def\kapustaolive{J.I. Kapusta and K. A. Olive, {\NPA{408}{478}{1983}}.}
\def\rischkegorenstein{.D.H. Rischke, M.I. Gorenstein, H. Stoecker and W. Greiner, Z. Phys. C {\bf 51}, 485 (1991).}
\def\moore{G. D. Moore and O. Saremi, JHEP {\bf 09}, 015 (2008).}
\def\greinerprl{J. Noronha-Hostler,J. Noronha and C. Greiner, Phys. Rev. Lett. {\bf 103}, (172302) (2009).}
\def\greinerprc{J. Noronha-Hostler, J. Noronha and C. Greiner, Phys. Rev C {\bf 86}, 024913 (2012).}
\def\gorenstein{M. Gorenstein, M. Hauer, O. Moroz, Phys. Rev. C {\bf 77}, 024911 (2008).}
\def\kubo{R. Kubo, J. Phys. Soc. Jpn. {\bf 12} (1957).}
\def\endrodi{G. Endrodi, JHEP {\bf 023}, 1304 (2013). }
\def\nam{Seung-il Nam, Chung-Wen Kao, Phys. Rev. D {\bf 87}  11, 114003 (2013).}
\def\itakura{K. Itakura, O. Morimatsu, and H. Otomo, Phys. Rev. D {\bf 77}, 014014 (2008).}
\def\cpsingh{S.K. Tiwari, P.K. Srivastava, C.P. Singh, Phys. Rev. C {\bf 85}, 014908 (2012).}
\def\guru{Guru Prakash Kadam, H. Mishra, arXiv:1408.6329.}
\def\rom{P. Romatschke, D. T. Son, Phys. Rev. D {\bf 80}, 065021 (2009).}
\def\ferrer{E.J. Ferrer, V. Incera, and J.P. Keith, I. Portillo, and P.
Springsteen, Phys. Rev. C {\bf 82}, 065802 (2010).}
\def\FukuKharzeev{K. Fukushima, Dmitri Kharzeev, and H. J. Warringa, Phys. Rev. D {\bf 78}, 074033 (2008).}

\end{document}